# Phase separation of metastable CoCrFeNi high entropy alloy at intermediate temperatures

Feng He[a], Zhijun Wang[a, *], Qingfeng Wu[a], Junjie Li[a], Jincheng Wang[a, *], C.T. Liu[b]

[a] *State Key Laboratory of Solidification Processing, Northwestern Polytechnical University, Xi'an 710072, China*

[b] *Center of Advanced Structural Materials, Department of Mechanical and Biomedical Engineering, College of Science and Engineering, City University of Hong Kong, Kowloon, Hong Kong, China*

**Abstract:** The CoCrFeNi alloy is widely accepted as an exemplary stable base for high entropy alloys (HEAs). Although various investigations prove it to be stable solid solution, its phase stability is still suspicious. Here, we identified that the CoCrFeNi HEA was thermally metastable at intermediate temperatures, and composition decomposition occurred after annealed at 750°C for 800 hrs. The increased lattice distortion induced by minor addition of Al into the CoCrFeNi base accelerated the composition decomposition and a second fcc phase with a different lattice constant occurred in the long time annealed CoCrFeNiAl0.1 HEA. A Cr-rich σ phase also precipitated from the CoCrFeNiAl0.1 HEA. The Al element can induce the instability of CoCrFeNi HEA. The revealed metastable CoCrFeNi at intermediate temperatures will greatly change the way of HEAs development.

**Introduction:**

The equal-atomic alloy of CoCrFeNi is widely accepted as a thermally stable single-phase solid-solution HEA for its high mixing entropy [1-11]. The CoCrFeNi HEA has face centered cubic (fcc) crystal structure and high toughness even at cryogenic temperatures [12-14]. However, its strength is far from industrial applications

* Corresponding author. zhjwang@nwpu.edu.cn
* Corresponding author. Tel:86-29-88460650; fax:86-29-88491484, jchwang@nwpu.edu.cn

[15], and consequently many researchers treated it as a base alloy. In history, the gradually developed single based alloys have greatly enhanced the civilization, such as the alloys of copper, iron, aluminum and nickel. After years of development, the mechanical properties of conventional alloys with a single base have almost reached their limits. Different from traditional single-based alloys, the CoCrFeNi alloy, as a high entropy base of a new catalog of alloys, has aroused much attention since the concept of HEAs was proposed [16-18]. On the one hand, the equal-atomic CoCrFeNi base is strengthened through solid solution by the mixing of alloying elements with different atomic size. On the other hand, the thermodynamic and kinetic behaviors of the single-phase solid-solution CoCrFeNi alloy will extend the current theory of physical metallurgy.

Usually, metallic materials are developed by adding alloying elements to make the matrixes stronger, tougher and stretchier. In the past decade, many researchers took the CoCrFeNi HEA as a base alloy and designed lots of new alloys by adding alloying elements [6, 7, 19-25]. He et al. [19] found that the CoCrFeNi HEA alloy can be strengthened from 400 MPa to 1300 MPa by adding Ti and Al elements. By adding Al element into the CoCrFeNi base, Ma et al. obtained fcc single crystals of $CoCrFeNiAl_{0.3}$ HEA with an ultimate tensile elongation of 80% [26]. Moreover, elements with larger atomic size were also used to strengthen the CoCrFeNi base and many new alloys were designed, such as $CoCrFeNiNb_x$ [22-24], $CoCrFeNiTi_x$ [21] and $CoCrFeNiMo_x$ [7]. These CoCrFeNi based HEAs are potential candidates to be applied in different environments.

Considering the potential applications of CoCrFeNi based alloys, the physical properties of CoCrFeNi matrix are of great significance from both scientific and industrial aspects. Lots of efforts have been made to reveal the phase stability of CoCrFeNi HEA. With XRD and neutron diffraction measurements, Lucas et al. [4] found that there is no chemical ordering in the CoCrFeNi HEA. Through TEM and atom probe tomography (APT) characterizations, Cornide et al. [11] also suggested one single fcc solid-solution phase in CoCrFeNi HEA. It was also found that the single fcc solid solution phase in CoCrFeNi was thermally stable after annealed at different temperatures [3-11, 27]. Besides the experimental results, all the present phase selection models, CALPHAD and first principle calculations indicated that the CoCrFeNi HEA should be a stable single fcc solid-solution phase [28-32]. Moreover, plenty of investigations also indicate that minor additions of alloying elements would not affect the phase stability of the CoCrFeNi base [6, 7, 31, 33, 34].

As one of the foundations of HEAs, the stability of the solid-solution base greatly affects the design of HEAs. Although all previous results have revealed that the CoCrFeNi base is stable, the very recent evidence shows that the CoCrFeMnNi HEA is unstable at intermediate temperatures [35, 36], which brings great challenges to the HEA community. Almost at the same time, by using XRD and neutron diffraction, Dahlborg et al. [37] claimed that the CoCrFeNi alloy should have two fcc phases with a discrepancy of 0.001Å in lattice parameter. These clues suggest the instability or meta-stability of the CoCrFeNi base. Considering the wide acceptance of the stable CoCrFeNi and CoCrFeNi based HEAs, it is intriguing and urgent to demonstrate whether the CoCrFeNi base is stable or not.

At high temperatures, the high mixing entropy effect [36] ensures the stability of solid solution of CoCrFeNi, and at low temperatures, the solid solution phase in the as-cast state is also stable due to the sluggish diffusion effect. The most possible instability case for the CoCrFeNi solid solution should be in the intermediate temperature range at its equilibrium state. Accordingly, in this paper, we reported the metastable state of CoCrFeNi at intermediate temperatures. The possible separation of composition is identified, and the decomposition of the matrix into two phases is confirmed with minor additions of Al element.

**Results and discussions:**

Fig. 1 shows the SEM image, EDX maps and XRD patterns of the long time annealed CoCrFeNi HEA. In the SEM image, a single phase HEA with grain boundaries was observed. Second phase was not found in the whole sample, neither in grain boundaries nor inside the coarse grains. The XRD patterns also showed regular single fcc peaks. The EDX maps confirmed that there was no composition decomposition in the CoCrFeNi HEA in the present resolution. It seems that these preliminary results in Fig. 1 agreed well with previous reports, where the CoCrFeNi HEA was identified to be thermally stable by TEM, APT, high energy x-ray scattering and neutron scattering [4, 11].

However, the details in TEM analyses showed distinct evidences of instability in the prolonged annealed CoCrFeNi HEA. The TEM bright field image is presented in Fig. 2(a). A large number of regions with darker fringes were observed. The insets in Fig. 2(a) show selected diffraction patterns of the matrix (inset A) and darker fringes (inset B) regions. The diffraction pattern of the matrix from [001] zone axis shows regular fcc crystal structure. In the darker region, two diffraction patterns from [011] zone axis are presented (marked as red and yellow), indicating a tiny difference of lattice parameters between the two crystals in the selected area. The red diffraction pattern could be identified as the matrix by comparing with the diffraction pattern of the matrix, while the yellow one should be a new separated phase. The elongated diffraction point of the new phase indicates that only very thin slice of new phase decomposed from the matrix.

To further figure out the phase separation, the lattice parameters of the two regions were separately characterized. Fig. 2(b) is the HRTEM image of the darker region and the inset is the corresponding fast Fourier transform (FFT) image. The FFT image shows that the darker region has the same crystal structure with the matrix. A typical atom arrangement with [001] zone axis of fcc single crystal is observed. The lattice parameter of the darker region is 0.355 nm measured from HRTEM image, a little bit different from that of the matrix (0.366 nm measured from diffraction pattern). Even though the bright field TEM image shows a different region with darker contrast, the HRTEM analysis did not found obvious interface of phase boundary across the two different regions and only lattice distortions were observed. Concerning the tiny difference of 0.01nm in lattice constant between the two phases and the absence of a clear phase interface, the dark fringes in the long time annealed CoCrFeNi HEA can be seen as the composition decomposition, the initial stage of phase separation. The decomposed composition contains atoms with different radii, inducing a difference in lattice constant.

The composition decomposition in the CoCrFeNi annealed at 750°C is similar to the clustering of GP zones. Usually, precipitates pass through several stages before a final stable structure appears. For example, the first stage in Al-4%Cu alloys involves local clustering of the solute atoms to form what are commonly called GP zones [38]. GP zones are favored at a low aging temperature, a small atomic size misfit, and a high degree of solute supersaturation. In the CoCrFeNi HEA, the atomic size misfit is small since there is only a small atom size difference among those component elements, but every element could be regarded as supersaturated solute atoms. During the long-term annealing at 750°C, the supersaturated CoCrFeNi solid solution may decompose into two phases. Besides, Dahlborg et al. [37] reported that the as-cast CoCrFeNi alloy consists two fcc phases with a lattice constant difference of 0.001Å using diffraction ways. The lattice constant difference in this study is much larger, so it is possible that the two fcc phases in the CoCrFeNi HEA evolved from metastable state to steady state. According to the above results, it is reasonable to deduce that a fcc phase with different lattice parameters can appear in the CoCrFeNi HEA through long time annealing at 750°C.

We have confirmed that the single solid solution CoCrFeNi HEA is metastable at 750°C. However, the long-term annealing is necessary to observe the phase separation. The long-term existence of metastable state is attributed to the large energy barrier or the small driving force for it to evolve to a steady state. From another perspective, increasing the lattice distortion may aggravate the composition decomposition. Moreover, Xu et al. [39] found that nano-scale phase separation can take place to efficiently minimize the lattice distortion caused by atom size difference of the constituent elements. Therefore, minor additions of alloying elements with large atomic radius will greatly accelerate the decomposition process. Here, the stability of CoCrFeNiAl0.1 was investigated to confirm this prediction. The limit solubility of Al element in Co, Cr, Fe and Ni at 750°C are much larger than 3 at.% according to their binary phase diagrams [40], thus minor addition of Al element does not lead to the formation of aluminide. However, the atom radius of Al element is much larger than that of other four elements. Consequently, the addition of Al element would lead to more severe lattice distortions. The XRD pattern in Fig. 1 showed that the addition of Al element made the peaks shift to left when compared with the CoCrFeNi HEA, which means that the lattice distortion of CoCrFeNiAl0.1 HEA increased. The larger lattice distortion caused by Al additions will lead to the instability of the CoCrFeNi HEA, differing from the reported conclusions [6, 20].

Fig. 3(a) shows the bright field TEM image of the prolonged annealed CoCrFeNiAl0.1 HEA. In Fig. 3(a), the darker regions were also found in CoCrFeNiAl0.1 matrix and the corresponding HRTEM image was shown in Fig. 3(b). Compared with the HRTEM of CoCrFeNi HEA, obvious Morie pattern can be seen in the HRTEM image of CoCrFeNiAl0.1 HEA, indicating distinctly two different phases with similar lattice parameters or two different grains. The FFT image in Fig. 3(b) showed that there were two different phases in the darker region. The lattice constants of the two phases are 0.424 nm and 0.365 nm measured from the FFT image, respectively.

According to the parallel Morie pattern theory, the Morie pattern period is calculated as $D = d_1 d_1 / (d_1 - d_2)$, where $d_1$ and $d_2$ are interplanar spacing of the two phases. The calculated period is $D = 0.212 \times 0.1825 / (0.212 - 0.1825) = 1.312$ nm. The period of the Morie pattern (D) was measured as 1.355 nm, agreeing well with the calculation. Morie pattern is always a powerful and accurate evidence of interplanar spacing difference between two different phases, so it is reasonable to claim that phase separated in the CoCrFeNi based HEA with minor additions of Al element. The reason for phase separation might be that the addition of Al element lead to more severe lattice distortions and thus the driving force of composition decomposition increased. The increased driving force accelerated the phase separation of the metastable CoCrFeNi HEA. The results identified the assumption of phase separation and further confirmed that the CoCrFeNi HEA is thermodynamic metastable at intermediate temperatures.

Besides the phase separation, large-scale precipitates were also observed in the CoCrFeNi base after the minor additions of Al element. Fig. 4 shows that a second phase precipitated from the grain boundary of the long time annealed CoCrFeNiAl0.1

HEA. The EDX maps in Fig. 4 show that the precipitate was enriched in Cr but poor in all other four elements. Chemical compositions from SEM-EDX in Table 1 present that the precipitate consists of at.71% Cr element, and also contained some Co, Fe and Ni, but surprisingly none of Al. Nevertheless, the XRD pattern of CoCrFeNiAl0.1 alloys in Fig. 1 did not show the evidence of precipitation due to the minor amount of the precipitate. The chemical composition of the Cr-rich phase and the corresponding annealing temperature were highly similar with that of the σ phase found by Schuh [41] and Pickering[35] in CoCrFeMnNi HEAs. Accordingly, it is reasonable to say that the Cr-rich phase found in this work is σ phase, as reported by Pickering et al.[35].

The appearance of Cr-rich σ phase in CoCrFeMnNi alloys is due to the rapid diffusion of Cr element at intermediate temperatures. Another reason for the formation of σ phase is that, the Cr element is a strong σ phase former which can be proved by that all the binary phase diagrams of Cr-Co, Cr-Ni and Co-Fe have σ phase zones [36]. In the CoCrFeNi system, the Cr element also has the largest diffusion coefficient and large σ phase forming ability [42], hence Cr-rich σ phase may also occur in the grain boundary. However, Cr-rich σ phase is not observed in the pure CoCrFeNi HEA, but only appears in the CoCrFeNiAl0.1 alloy after long-term annealing at 750°C. A possible reason of this difference is attributed to the lattice distortion induced by Al addition. Zhang et al. [43] reported that severe lattice distortions would lead to phase transformation. Therefore, addition of Al element induced the instability of the fcc CoCrFeNi matrix for the increased distortion energy.

**Conclusions**

In summary, the phase separation of CoCrFeNi and CoCrFeNiAl0.1 HEAs in intermediate temperature was confirmed by using SEM-EDX, XRD and TEM for the first time. The results revealed that the CoCrFeNi HEA was thermally metastable through 800 hrs-annealing, and with a minor addition of Al element it became unstable. Composition decomposition occurred in the CoCrFeNi HEA after prolonged exposure at 750°C. The addition of Al element aggravated the decomposition, resulting in the phase separation. The new phase is also a fcc phase but with a different lattice constant. Besides, a Cr-rich σ phase also precipitated from the CoCrFeNiAl0.1 HEA. The occurrence of precipitates in the CoCrFeNiAl0.1 HEA confirmed that the addition of Al element induced the instability of metastable CoCrFeNi HEA. As a result, the phase stability of CoCrFeNi and CoCrFeNiAl0.1 HEAs should be reconsidered and the use of existing phase selection models should be limited to special environments, e.g., at high temperatures.

**Experimental**

HEAs ingots were prepared by arc-melting under an argon atmosphere. The purity of raw materials is 99.5%. To achieve a homogeneous distribution of elements in the alloys, each ingot was re-melted for four times. Heat-treatment of the HEA plate samples were performed under air atmosphere at 750 ºC for 800 hrs followed by water-quenching. The phases were characterized by an X-ray diffractometer (XRD, Bruker D8 discover), using Co Kα radiation scanning from 20º to 120º with a scanning rate of 1º/s. The scanning electron microscope (TESCAN VEGA 3) equipped with energy dispersive X-ray spectrometry (EDX) was used to analyze the microstructures and chemical compositions. Standard bright-field images and diffraction patterns were obtained using a transmission electron microscope (TEM TecnaiFG2).

**Acknowledgements**

The work was supported by National Natural Science foundation of China (Grant No. 5147113).

**Author contributions**

Zhijun Wang，Jincheng Wang and Feng He proposed the idea. The experiments were performed by Feng He and Qingfeng Wu. Junjie Li and C.T. Liu contribute to the results analysis and discussion. Feng He and Zhijun Wang wrote the paper.

**Additional Information**

Competing financial interests: The authors declare no competing financial interests

Table 1. Chemical compositions of precipitates measured by SEM-EDX (at.%).

| Precipitates | Co | Cr | Fe | Ni | Al | Notes |
|---|---|---|---|---|---|---|
| σ phase | 8.84 | 71.24 | 10.20 | 9.73 | 0 | SEM-EDX |

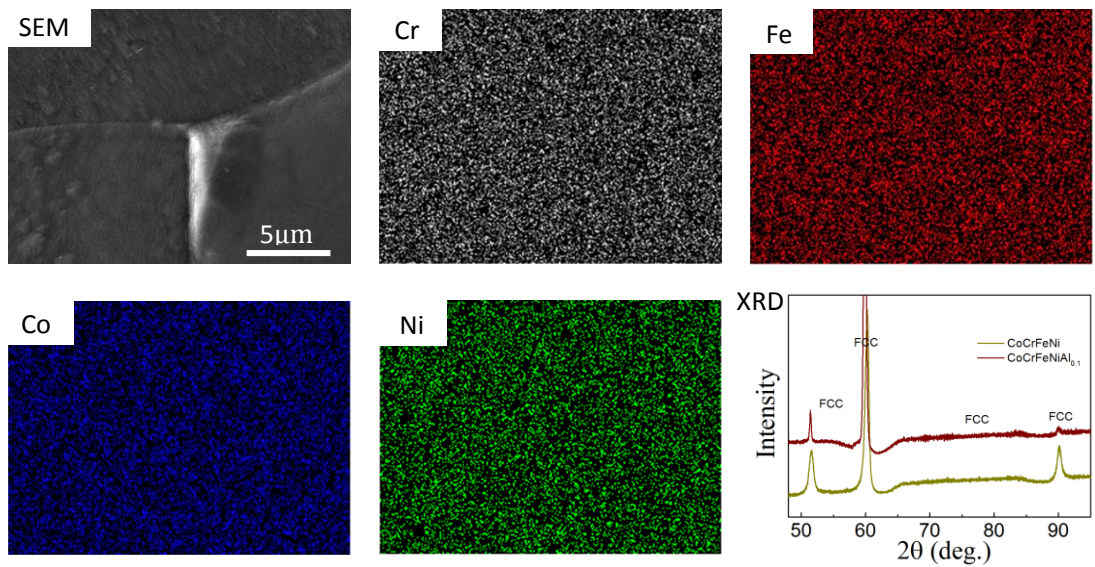

Fig. 1. SEM images and EDX maps of the CoCrFeNi HEA annealed at 750°C for 800 h. EDX maps illustrated that there is no composition decomposition in the long time annealed CoCrFeNi HEA. The XRD patterns of the CoCrFeNi and CoCrFeNiAl0.1 alloys showed that both the two alloys have fcc crystal structures and the CoCrFeNiAl0.1 HEA has a relative larger distortion.

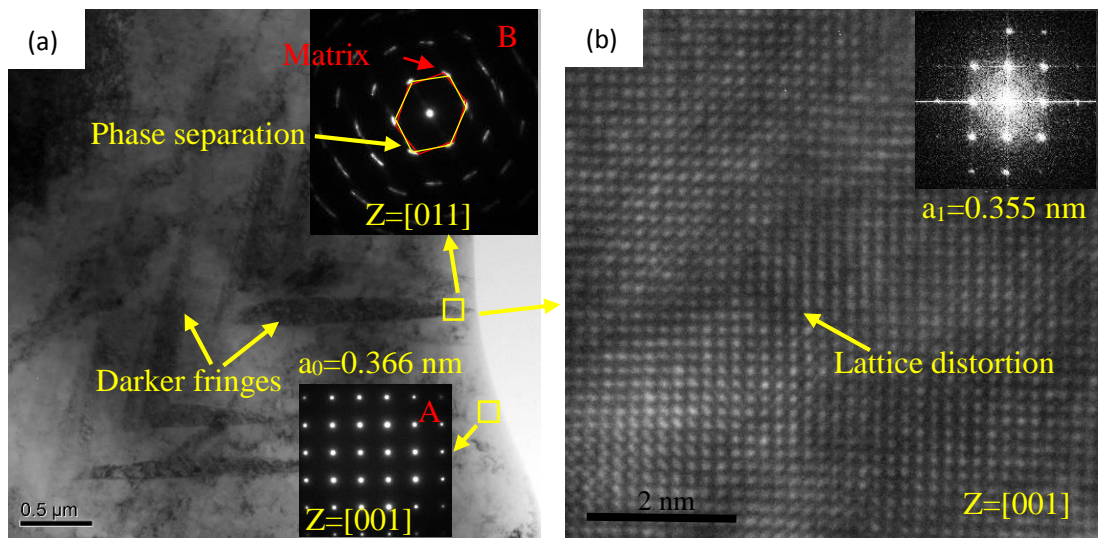

Fig. 3. The (a) bright field TEM and (b) HRTEM images of the long time annealed CoCrFeNi HEA. The insets of (a) are selected diffration patterns of A darker fringes area and B the matrix, respectively. The lattice constant of the matrix measured from B is $a_0 = 0.366$ nm. The inset of (b) is the FFT image of (b) and the corresponding measured lattice constant is $a_1 = 0.355$ nm.

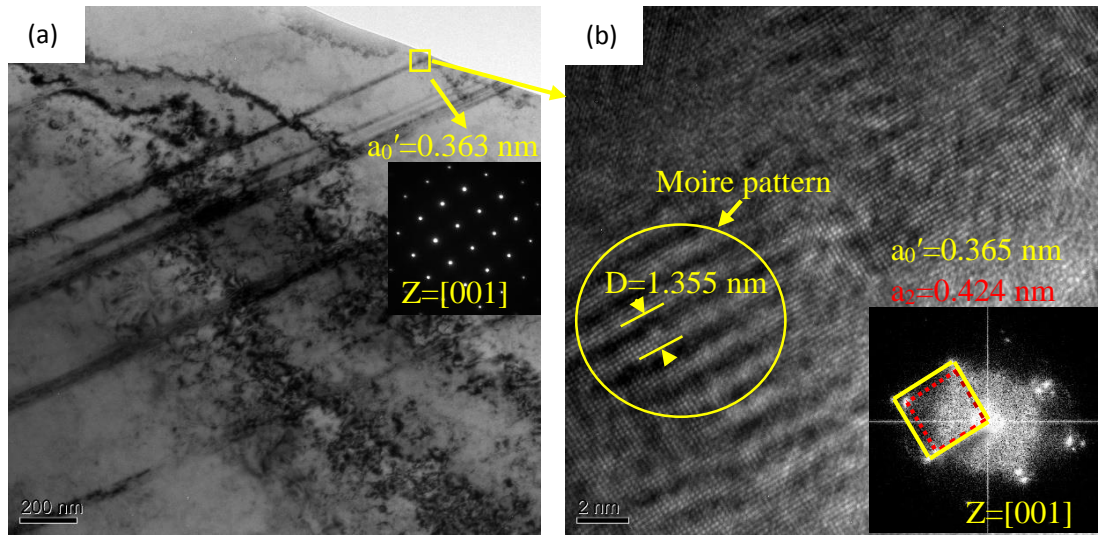

Fig. 3. TEM bright field images of (a) CoCrFeNiAl0.1 HEAs, and the insets are selected diffraction patterns of squares. The lattice constant of the CoCrFeNiAl0.1 matrix measured from the diffraction patterns is $a_0' = 0.363$ nm; (b) is HRTEM image of the squares in (a), the insets are reduced FTT images of (b); the lattice constant of the matrix and second phase are $a_0' = 0.365$ nm and $a_2 = 0.424$ nm. The period of the Moire pattern was measured to be $D = 1.355$ nm.

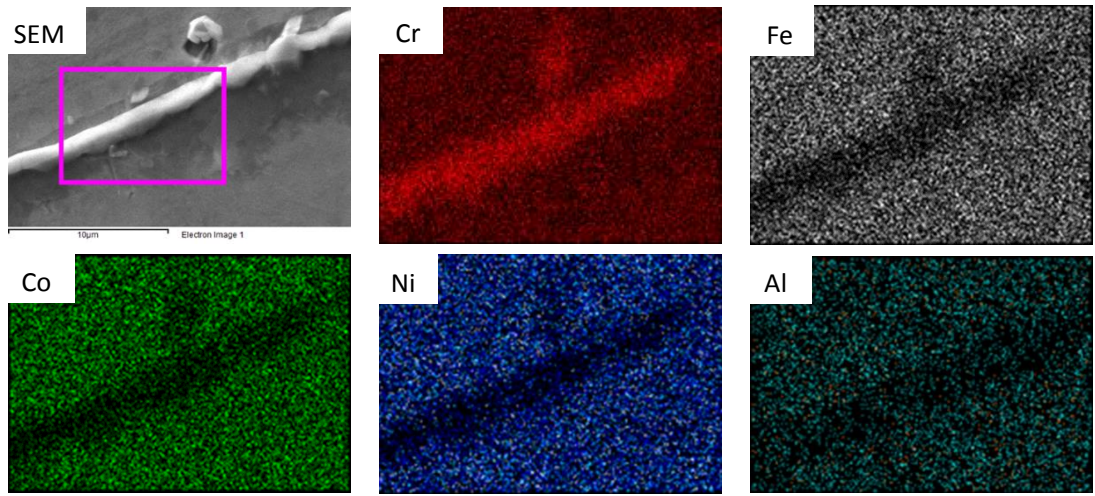

Fig. 4. SEM images and EDX maps of CoCrFeNiAl0.1 HEAs annealed at 750ºC for 800 h. The EDX maps indicated that the precipitate is rich in Cr element and rare in all other four components, including Al.